\documentclass[runningheads]{llncs}

\usepackage[utf8]{inputenc}

\usepackage{algorithm}
\usepackage[noend]{algorithmic}
\usepackage{amsmath,amssymb,amsfonts}
\usepackage{stmaryrd}
\usepackage{xcolor}
\usepackage{comment}
\usepackage{url}
\usepackage{multirow}
\usepackage{graphicx}
\usepackage{booktabs}

\usepackage{appendix}

\DeclareMathOperator*{\RandMech}{\mathcal{M}}

\newcommand{\aux}{aux}

\newcommand{\quantPoissonFunc}[1]{Q_{#1}}
\newcommand{\quantPoisson}[2]{Q_{#1}\left(#2\right)}

\newcommand{\nBits}{nBits}

\newtheorem{observation}{Observation}

\begin{document}
\title{Protecting Data from all Parties: Combining Homomorphic Encryption and Differential Privacy in Federated Learning}
\titlerunning{Protecting Data from all Parties in FL}

\author{
    Arnaud Grivet S\'{e}bert\inst{1}\and
    Renaud Sirdey\inst{1}\and
    Oana Stan\inst{1}\and
    C\'{e}dric Gouy-Pailler\inst{1}
}
\authorrunning{A. Grivet Sébert et al.}

\institute{Institut LIST, CEA, Université Paris-Saclay,\\F-91120, Palaiseau, France\\
\email{\{arnaud.grivetsebert,renaud.sirdey,oana.stan,cedric.gouy-pailler\}@cea.fr}}
\maketitle
\begin{abstract}
This paper tackles the problem of ensuring training data privacy in a federated learning context. Relying on Homomorphic Encryption (HE) and Differential Privacy (DP), we propose a framework addressing threats on the privacy of the training data. Notably, the proposed framework ensures the privacy of the training data from all actors of the learning process, namely the data owners and the aggregating server. More precisely, while HE blinds a semi-honest server during the learning protocol, DP protects the data from semi-honest clients participating in the training process as well as end-users with black-box or white-box access to the trained model.
In order to achieve this, we provide new theoretical and practical results to allow these techniques to be rigorously combined. In particular, by means of a novel stochastic quantisation operator, we prove DP guarantees in a context where the noise is quantised and bounded due to the use of HE.
The paper is concluded by experiments which show the practicality of the entire framework in terms of both model quality (impacted by DP) and computational overhead (impacted by HE).

\keywords{Federated learning \and Differential privacy \and Homomorphic encryption \and Quantisation.}
\end{abstract}

\section{Introduction}
Machine learning techniques have become ubiquitous in almost all fields of industry and our daily lives. The used algorithms, and especially neural networks, the most popular ones in many applications, require massive amounts of data to get trained and reach the impressive accuracy that made their success. This huge need for data together with the omnipresence of the Internet and the boom of communication exchanges made data the "new oil" (Clive Humby, 2006).

In parallel, confidentiality has raised greater and greater concerns as evidenced by the new regulations on data privacy (e.g. GDPR~\cite{EUdataregulations2018} in the EU) and, while all kinds of data might be considered as private, some fields, like healthcare (e.g. HIPAA \cite{hipaa} in the USA) or military applications, are especially sensitive to privacy breaches.
In this context, it is well known that, once trained, a machine learning model may indirectly release some information about its training data \cite{hannun2021measuring}. Many researchers have exhibited attacks on these models, assuming that the adversary has access to the parameters of the model (white-box access) or even only to the output of the inference on some queries she made to the model (black-box access) \cite{shokri2017membership,sablayrolles2019white,fredrikson2015model}. These attacks are even more likely with the emergence of machine learning as a service~\cite{ribeiro2015mlaas} in the recent years.

To deal with these issues, the most popular approach to data privacy is differential privacy (DP), introduced by Dwork et al.~\cite{dwork2006our}, which quantifies the amount of information leaked by the output of a mechanism about its input. This notion of privacy implies that the considered mechanism is probabilistic, which is generally achieved by applying carefully parameterised random noise to a deterministic mechanism. Since 2006, DP has been widely used in machine learning applications (see e.g. \cite{ji2014differential}).

Cryptography is another field that helps providing protection to sensitive data, in particular at training time. Specifically, homomorphic encryption (HE) \cite{Phong2017,yang2019federated} or multi-party computation \cite{Bonawitz2017,Mugunthan2019} allow to perform secure, blind computations without access to the inputs, and, in the case of HE, the outputs either. Additionally, HE can in some cases be associated to tools such as verifiable computing to bring additional computation integrity guarantees, including in the context of federated learning \cite{madi2021secure}.

In 2016, McMahan et al. proposed a new paradigm of collaborative learning that they called federated learning (FL) \cite{mcmahan2016federated}. Along with the reduction of communication load and the parallelism it allows, a claimed key advantage is the protection of data due to the fact that each client keeps its own data locally. However, although FL gives some protection to the data with regard to the server, it gives rise to a new type of potential adversaries - the other clients. Several attacks that take advantage of this new threat were proposed \cite{hitaj2017deep,melis2019exploiting}.

The contribution of this paper is the design of a privacy-preserving FL framework consistently combining countermeasures of different natures and aims, namely DP and HE. The above-mentioned attacks on the training data are addressed via DP, either they come from participants during the training or from the end-users of the model. Another potential threat comes from the server. To hide the data from it, the clients send encrypted information to the server which will do the necessary computations in the encrypted domain, without seeing either the sent information or the result of its computations, thanks to the homomorphic properties of the cryptosystem. 
In order to consistently articulate DP and HE we introduce a new stochastic quantisation operator based on Poisson distribution. This operator behaves as it was applied as post-processing of a Gaussian mechanism, keeping the DP guarantees of this standard mechanism unchanged without any supplementary analysis and allowing to seamlessly get rid of the quantisation issue due to the use of HE.

An interesting application scenario concerns the medical field. We may consider several hospitals that own medical data from their patients and wish to collaborate in order to train a global model that would detect a certain disease. In many countries, patient data are sensitive and the hospitals do not want to share them with other hospitals. A solution is to use an aggregation server (e.g. from an institutional entity) but the hospitals do not want to give it their data either.
Note that the parameters we used for our experiments on FEMNIST (see Section~\ref{sec:experiments}) make our cross-silo solution scales at a level which can be for example compatible with the number of medical facilities in a reasonably large country.

The paper is organised as follows. In Section~\ref{sec:related_work}, a review of the literature on the issues of data privacy in FL context follows this introduction. Then, Section~\ref{sec:preliminaries} provides the technical prerequisites necessary to understand our method, that we thoroughly explain in Section~\ref{sec:method}. The results of the experiments that we ran to illustrate the feasibility of our solution are presented in Section~\ref{sec:experiments}, before conluding remarks and perspectives for further work (Section~\ref{sec:conclusion}).

\section{Related work}
\label{sec:related_work}

\subsection{Differentially private federated learning}
Due to the additional threats from the other clients, DP has been quite popular in collaborative and especially FL frameworks.
For instance, McMahan et al.~\cite{mcmahan2017learning} trained a next word prediction algorithm in a federated way, using text data from users' smartphones, protected by DP. Nevertheless, this work is hardly comparable to ours because the task for which the model is trained is very different from our targeted classification tasks (e.g. FEMNIST in our experiments), inducing very different learning parameters and thus a quite different privacy challenge.
Closer to our work is the one from \cite{geyer2017differentially} which presents a differentially private FL process tested on MNIST, yet a notably easier task than FEMNIST.
Shokri et al.~\cite{shokri2015privacy} also experimented a differentially private distributed learning setting - similar to a FL setting due to the use of distributed stochastic gradient descent - on MNIST, and SVHN, another famous image dataset.
In~\cite{sabater2020distributed}, Sabater et al. ensures privacy and utility of their distributed learning setting as long as each participant communicates with only a logarithmic number of other participants.
Abadi et al.~\cite{abadi2016deep} introduced the moments accountant method, a useful tool for our DP analysis, that enables to keep track of the privacy cost more tightly than the traditional composition theorem when many calls to the database are done, typically for training a deep neural network.

\subsection{Cryptographic primitives for federated learning}
\label{sec:related_crypto}
Most of the works applying HE to machine learning models focus on the inference stage (CryptoNets \cite{cryptonet}, TAPAS \cite{Sanyal2018}, NED \cite{Hesamifard2019}) and not on the training stage.
The first papers on privacy-preserving machine learning training focused on a centralised setting where all data are outsourced and where the models are only linear \cite{bergamaschi2019homomorphic,Carpov2019}. When it comes to non-linear models, the few approaches that ran a complete centralised training of neural networks on encrypted data have impractical performances or huge cryptographic parameters \cite{lou2019glyph}.

While some authors propose solutions in the case of multi-servers either for clustering or regression, many methods employing HE have been recently proposed for a collaborative learning task with no central server. They mostly apply on linear models \cite{zheng2019helen,li2020faster} and, more recently, Sav et al. focused on neural networks \cite{sav2020poseidon}.
As far as FL is concerned, only a few recent papers propose the use of HE to protect the clients' data from the server \cite{Phong2017,yang2019federated,zhang2020}, \cite{Phong2017} and \cite{yang2019federated} being only theoretical.

More general than HE, multi-party computation protocols like secure aggregation allow several agents to collaborate and compute a function on their data such that each agent knows no more than its own input and, if requested, the output, and learns nothing about the other agents' inputs. The combination of the high-communication costs of the multi-party computation and the inherent distributed nature of FL makes FL methods secured by multi-party computation (e.g. \cite{Bonawitz2017,Mugunthan2019}) difficult to implement efficiently. Among these approaches, \cite{jayaraman2018distributed} interestingly makes use of DP techniques to further protect the private data.

\subsection{Quantisation and differential privacy}
\label{sec:related_quantisation}
A great focus and key contribution of our paper deals with the interference between our two defensive tools, namely DP and HE. The main issue induced by this interference is that the range of the messages to be encrypted (which are the noised updates in our work) has to be discrete and bounded (and representable with as few bits as possible).
For potentially other reasons than a necessary encryption of the data, some authors have studied the possibility of making a machine learning mechanism differentially private with a discrete noise.

The authors of \cite{agarwal2018cpsgd} propose a secure and communication-efficient distributed learning framework and perform the DP analysis of their learning mechanism using a binomial noise because the effect of quantisation on the Gaussian mechanism is unclear, especially after aggregation if the noise is generated in a distributed way. The analysis is quite involved and only provides DP bounds for the multidimensional binomial mechanism for one round of learning. Indeed, the moments accountant method is not easily applicable to the binomial distribution. Moreover, the presented DP guarantee is worse than the Gaussian mechanism's one and needs the quantisation scale to tend to zero (and the communication cost to infinity) to approach it.

In \cite{koskela2021tight}, Koskela et al. present a privacy accountant for discrete-valued mechanisms for non-adaptive queries using privacy loss distribution formalism and Fast Fourier Transform. In particular, they give DP guarantees for the binomial mechanism in one dimension and extend them in the multidimensional case but with quite demanding constraints that compel them to brutally approximate the gradients by their sign in their experiments.
Cannone et al.~\cite{canonne2020discrete} introduce the discrete Gaussian mechanism and studied its DP guarantees that scale well with composition, even in the multivariate case. Nevertheless, contrary to binomial noise, discrete Gaussian noise is not bounded as required for our framework. Besides, the discrete Gaussian distribution is not stable by addition, thus precluding its direct use in a context of distributively generated noise that a collaborative learning task with untrusted server requires (see Section~\ref{sec:distributed_noise}).

In \cite{kairouz2021distributed,agarwal2021skellam}, the authors propose federated learning protocols protected by DP and secure aggregation (which requires discrete and bounded values, as HE, but needs communication before learning as mentioned in~\ref{sec:related_crypto}). These works respectively use the discrete Gaussian mechanism and the Skellam mechanism to ensure DP. At the cost of a careful DP analysis, they show that, for fine enough quantisation scale, their DP guarantees approach the Gaussian mechanism's one. Our work proposes a much simpler way to obtain the very same guarantee as the Gaussian mechanism, without needing to constrain the quantisation scale and with much simpler mathematical analysis. Moreover, the two previous works have to make use of conditional randomized rounding to ensure that the rounding of the unnoised values does not increase their norm too much. Since we perform quantisation after noising with a quantisation that can be viewed (from the DP perspective) as a post-processing (see Section~\ref{sec:poisson_quantisation}), we do not have such an issue.


\section{Preliminaries}
\label{sec:preliminaries}

\subsection{Federated learning}
FL is a decentralised framework that enables multiple agents, called \emph{clients}, to collaboratively train a shared global model under the orchestration of a central server while keeping the training data localised on the client devices thus helping to protect the data privacy and reducing communication costs. After a common (server-side) arbitrary initialisation of the global model, the FL process consists of successive rounds of communication between the server and the clients.

The most common approach to optimisation for FL is the Federated Averaging algorithm \cite{McMahan2017}. At the beginning of each round, the server selects a subset of clients to take part in training for this round, we call these particular clients the \emph{participants}. The server sends the current global model to the participants and each of them trains the model locally with several epochs of stochastic gradient descent (SGD) using its own data. The participants then communicate only the updated parameters or the updates\footnote{Difference between the updated parameters and the old ones.} back to the server. In our work, the participants send the updates because their norm is easier to constrain (necessary for privacy reasons). Finally, the server computes the weighted average of these updates before accumulating them into the global model, thereby concluding the round. The weight associated to a participant in the average is generally the fraction of training samples owned by the participant.
Throughout the paper, $M$ is the total number of clients; $K$ is the number of participants per round; the participants are indexed by $k$ with $n_k$ the number of training samples of $k$.

\subsection{Differential privacy}
\label{sec:preliminaries_dp}
DP~\cite{dwork2006our} is a gold standard concept in privacy-preserving data analysis. It provides a guarantee that, under a reasonable privacy cost $(\epsilon,\delta)$, two \emph{adjacent} databases produce statistically indistinguishable results. The notion of adjacence varies among authors. In our FL framework, the term database denotes the concatenation of all the clients' databases and two databases are adjacent if they have the same number of clients and differ on a single client, all the others remaining unchanged. Yet, the differing clients may have totally different data, making our notion of adjacence quite conservative (this is called \emph{user-level privacy}).
\begin{definition}
   Given $(\epsilon, \delta)\in \left(\mathbb{R^*_+}\right)^2$, a randomised mechanism $\RandMech$ with output range $\mathcal{R}$ satisfies \emph{$(\epsilon,\delta)$-DP} if, for any two adjacent databases $d,d'$ and for any subset $S\subset \mathcal{R}$, one has \begin{equation*}
       \mathbb{P}\left[ \RandMech(d) \in S\right] \leq e^\epsilon \mathbb{P}\left[\RandMech(d') \in S\right] +\delta.
   \end{equation*}
\end{definition}

One of the most widely used DP mechanism is the \emph{Gaussian mechanism}, which simply adds a Gaussian noise of mean $0$ and standard deviation $\sigma\in \mathbb{R^*_+}$, to each component of the output.

A key property of DP is its immunity to \emph{post-processing}, as stated below.

\begin{proposition}[\cite{dwork2014algorithmic}]
	\label{prop:post_processing}
	Let $\RandMech$ be a randomized	algorithm, with output range $\mathcal{R}$, that is $(\epsilon, \delta)$-differentially private, with $(\epsilon, \delta)\in \left(\mathbb{R^*_+}\right)^2$. Let $f \colon \mathcal{R} \to \mathcal{R'}$ be an arbitrary randomized mapping. Then $f \circ \RandMech$ is $(\epsilon, \delta)$-differentially private.
\end{proposition}

To determine the privacy cost $(\epsilon, \delta)$ of our protocol, we use a traditional two-fold approach. First of all, we determine the privacy cost per query and, in a second step, we compose the privacy costs of all queries to get the overall cost. Since training a deep neural network requires a large amount of calls to the database, we need to keep track of the privacy cost with a more refined tool than classical composition theorem (see e.g.~\cite{dwork2014algorithmic}), namely the \emph{moments accountant}~\cite{abadi2016deep}.

\begin{definition}
    The \emph{moments accountant} is defined for any $l\in \mathbb{N^*}$ as
    \begin{equation*}
        \alpha_{\RandMech}(l):=\max_{\aux,d,d'}\log \left(\mathbb{E}_{o \sim \RandMech(\aux,d)}\left[\left(\frac{\mathbb{P}[\RandMech(\aux,d) =o]}{\mathbb{P}[\RandMech(\aux,d') =o]}\right)^l\right]\right)
    \end{equation*}
    where the maximum is taken over any auxiliary input $\aux$ and any pair of adjacent databases $(d,d')$.
\end{definition}

\begin{theorem}[\cite{abadi2016deep}]
     \label{theorem:moments_composition}
      Let $p\in \mathbb{N}^*$. Let us consider a mechanism $\RandMech$ defined on a set $\mathcal{D}$  that consists of a sequence of adaptive mechanisms $\RandMech_1, \dots, \RandMech_p$ where, for any $i\in \{1, \dots, p\}$, $\RandMech_i\colon \prod_{j=1}^{i-1}\mathcal{R}_j\times \mathcal{D} \mapsto \mathcal{R}_i$. Then, for any $l\in \mathbb{N^*}$,
    \begin{align*}
        \alpha_{\RandMech}(l) \leq \sum_{i=1}^p \alpha_{\RandMech_i}(l).
    \end{align*}
\end{theorem}

Finally, once $\delta$ is chosen, the DP guarantee is derived from the overall moments accountant applying the tail bound property, stated in Theorem~\ref{th:tail_bound} from~\cite{abadi2016deep}.

\begin{theorem}[\cite{abadi2016deep}]
    \label{th:tail_bound}
    For any $\epsilon \in \mathbb{R}_+^*$, the mechanism $\RandMech$ is $(\epsilon, \delta)$-differentially private for $\delta = \min_{l\in \mathbb{N}^*} \exp(\alpha_{\RandMech}(l) - l\epsilon)$.
\end{theorem}


\section{Our privacy-preserving federated learning framework}
\label{sec:method}

\begin{figure}[htbp]
    \centering	\includegraphics[width=0.7\linewidth]{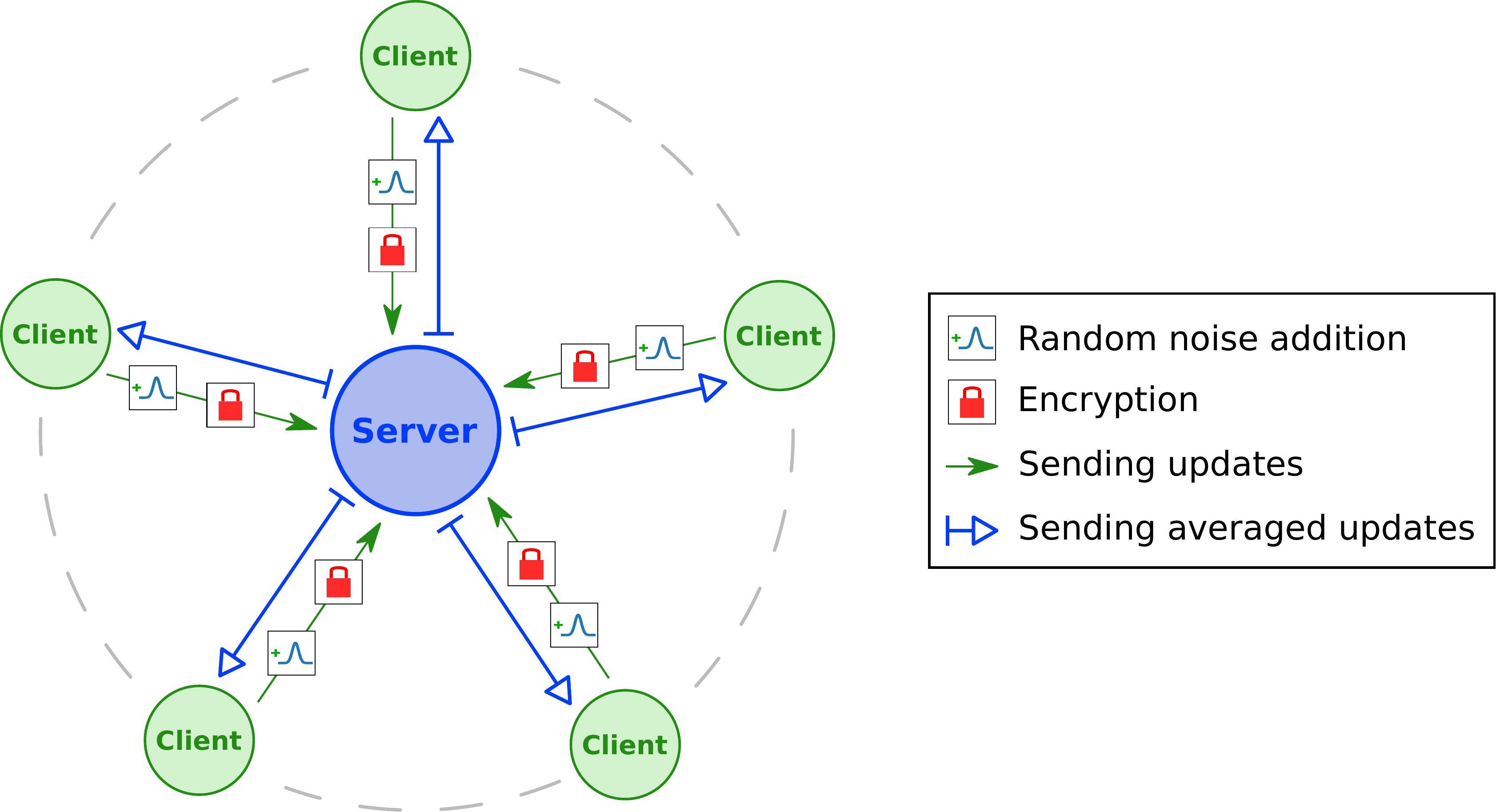}
    \caption{Our secure federated learning architecture.}
    \label{fig:architecture}
\end{figure}

Our FL framework is illustrated in Figure~\ref{fig:architecture}. After a common initialisation of the model, the process at each iteration is the following:
\begin{itemize}
    \item each participant performs a SGD of the loss function using its local data
    \item each participant applies on the obtained updates the successive transformations required by DP and HE (clipping, noising and quantisation)
    \item each participant encrypts and sends the transformed updates to the server
    \item the server aggregates (averages) the updates in the encrypted domain
    \item the server sends the current model parameters to the participants
    \item each participant decrypts the parameters thanks to the decryption key
\end{itemize}

\textit{\textbf{Threat model:}}
The adversaries that we consider are all the actors of the learning process, namely the clients and the server, and all the entities having access to the model once training is over, namely the end-users (that may include the clients themselves). The clients are assumed to be honest-but-curious, which means that they properly perform their task but may use the information they get to retrieve sensitive information about the training data. The end-users are also assumed to be semi-honest, although they do not necessarily participate to the training protocol. The server is also assumed to be mostly honest-but-curious and, as such, it is assumed to perform the correct required computations. However, as a slight bonus beyond this assumption, the server is not trusted to generate the noise needed for DP in order to prevent the threat whereby that noise is shared with a client or an end-user. As explained in the end of Section~\ref{sec:DP_analysis}, by easily adapting the DP guarantee, the same misbehaviour can also be taken into account for the clients that may collude and share their noises.

\subsection{Distributed noise generation}
\label{sec:distributed_noise}

When one wants to protect the training data by DP in a collaborative learning process, the assumption of the honest-but-curious server might be mitigated and the server may not be trusted to generate the noise, because it may communicate it to clients or end-users, thus annihilating the DP guarantees.
A common practice in that case is to make the participants generate the noise in a distributed way \cite{sabater2020distributed,grivet2021speed}.
This is especially practical when one wants the resulting noise to follow a Gaussian distribution, since this distribution is stable by addition. The participants simply need to generate Gaussian noise with well-chosen variance.
However, DP in a FL context still requires adaptations of the FL process:
\begin{itemize}
    \item clipping the updates in L2-norm with clipping bound $S$ (substituting participant $k$'s vector of updates $u_k$ by $\min\left(1, \frac{S}{\|u_k\|_2}\right).u_k$) to bound the sensitivity (i.e. the impact of changing from one dataset to an adjacent one) since unbounded sensitivity is incompatible with any DP guarantee
    \item adding noise to the gradients (e.g. Gaussian noise)
    \item fix all the coefficients of the mean to $\frac{1}{K}$, independently of the size of the participant's dataset, to bound the sensitivity more easily
\end{itemize}

\subsection{Problem of the limited number of bits and first approaches}
In our scenario, the information sent by the participants to the server is encrypted via HE. Since floating-point homomorphic calculations (although in principle possible) are prohibitively costly, we have to switch to a fixed-point representation while avoiding using too many bits (as the cost of FHE calculations will decrease, due to several factors discussed in Section \ref{sec:experiments}, with the number of bits required to represent the plaintexts). This means that, unlike the usual case where the noise is represented by a double-precision float (i.e. finite but very fine precision) and where we make the assumption that it perfectly follows the desired distribution, we here have to explicitly take into account bounds on the noise and quantisation of this noise (and of the updates themselves).
However, if we round the noised updates in a traditional way (scaling and rounding to the nearest integer), the aggregated noise is not Gaussian any more, but it is a sum of noises that follow rounded Gaussian distributions. Unfortunately, the distribution of a sum of rounded Gaussian variables does not have a simple expression (in particular, it is not a rounded Gaussian distribution)\footnote{If such a distribution would be easily dealt with, the issue of boundedness could be addressed by considering the modulo operation induced by encryption, like in~\ref{sec:unboundedness}.}.

A naive approach would be to try to compare the complicated distribution of a sum of rounded Gaussians with the one of a sum of perfect Gaussians i.e. a Gaussian. Indeed, if the quantisation scale is fine enough, the sum of rounded Gaussians should intuitively be close to a Gaussian. The final privacy cost should be the sum of the one of a classical Gaussian mechanism and a hopefully small additional privacy cost due to the approximation. Nevertheless, this analysis is quite involved and may result in overestimated DP bounds.

Another idea is to use binomial noise instead of Gaussian noise as in~\cite{agarwal2018cpsgd,koskela2021tight} but, as explained in Section~\ref{sec:related_quantisation}, the binomial distribution is not rotation invariant, which makes it hard to use in multidimensional problems and the moments accountant method does not apply directly to it, as it does for the Gaussian distribution. Besides, the DP guarantee obtained in~\cite{agarwal2018cpsgd} with an involved mathematical analysis needs the quantisation scale to tend to $0$ to get close to the Gaussian mechanism's guarantee while communication constraints precisely require this scale to be large.

We will now show that, under natural practical assumptions and with the use of a novel specific quantisation function, we may significantly simplify the DP analysis, with no privacy loss compared to the Gaussian mechanism. Most importantly, our DP guarantee does not depend on the quantisation scale, letting us free from the trade-off between privacy and communication faced in~\cite{agarwal2018cpsgd,kairouz2021distributed,agarwal2021skellam}.

\subsection{Poisson quantisation}
\label{sec:poisson_quantisation}
We here propose a new probabilistic quantisation operator that commutes with the sum\footnote{Commutativity must be understood in a large sense, as the offset parameter of the quantisation changes depending on the order of the operators.}, and is therefore harmless for the DP guarantee of the mechanism.
In the following, $\mathcal{P}(\lambda)$ denotes the Poisson law of parameter $\lambda \in \mathbb{R}^*_+$ whose support is $\mathbb{N}$ and whose probability mass function is $k\in \mathbb{N} \mapsto \frac{\lambda^k}{k!} e^{-\lambda}$.
We fix the quantisation scale $s \in \mathbb{R}^*_+$ and the dimension $d\in \mathbb{N}^*$ of the problem (the number of parameters of the model in our case).

\begin{definition}
	Let $\mu \in s\mathbb{Z}$.
	We define the probabilistic function
	\begin{align*}
	    \quantPoissonFunc{s, \mu} \colon x\in ]\mu; +\infty[ \mapsto s Y + \mu
	\end{align*}
	where $Y \sim \mathcal{P}\left(\frac{x-\mu}{s}\right)$. We call it the \emph{Poisson quantisation} of scale $s$ and offset $\mu$.
	
	Similarly, we define $\quantPoissonFunc{s, \mu} \colon x = \left(x^{(i)}\right)_{i\in \llbracket 1; d\rrbracket}\in ]\mu; +\infty[^d \mapsto \left(\quantPoisson{s, \mu}{x^{(i)}}\right)_{i\in \llbracket 1; d\rrbracket}$.
\end{definition}

Given $\mu \in s\mathbb{Z}$, for all $x\in ]\mu; +\infty[^d$, $\quantPoisson{s, \mu}{x} \in (s\mathbb{Z})^d$ and its mean is equal to $x$ so we can actually consider the Poisson quantisation as a function of probabilistic quantisation.
Proposition \ref{prop:quant_poisson_postproc} shows that the Poisson quantisation on the terms of a sum can be considered as a post-processing on the sum.

\begin{proposition}
	\label{prop:quant_poisson_postproc}
	Let $m\in \mathbb{N}^*$, $x_1, \dots, x_m \in \mathbb{R}$. Let $\mu\in s\mathbb{Z}$ such that $\mu < \min\{x_i | i\in \llbracket 1; m\rrbracket\}$.
	$\quantPoisson{s, m\mu}{\sum_i^m x_i}$ has the same distribution as $\sum_i^m \quantPoisson{s, \mu}{x_i}$.
\end{proposition}

\begin{proof}
	$\sum_i^m \quantPoisson{s, \mu}{x_i} \sim \sum_i^m (sY_i + \mu) = s\sum_i^m Y_i + m\mu$ where, for all $i\in \llbracket 1; m\rrbracket$, $Y_i \sim \mathcal{P}(\frac{x_i - \mu}{s})$. By stability of the Poisson law by addition, we know that $\sum_i^m Y_i \sim \mathcal{P}\left(\sum_i^m\frac{x_i - \mu}{s}\right) = \mathcal{P}\left(\frac{\sum_i^m x_i - m\mu}{s}\right)$. We then directly get the result.
\end{proof}

Proposition~\ref{prop:quant_poisson_postproc} together with Proposition~\ref{prop:post_processing} enables us to conclude that Poisson quantisation has no influence on the DP guarantee. Indeed, the output distribution is the same as if we had applied the Poisson quantisation after the aggregation of the \emph{continuously} noised updates. Since adding continuous Gaussian noises distributively on the updates and add them afterwards amounts to add a Gaussian noise to the sum of the unnoised updates, the Poisson quantisation acts as if it was applied on top of the Gaussian mechanism. \emph{Hence, the huge advantage of our Poisson quantisation operator is that it allows to reduce the DP analysis back to the vanilla analysis of the Gaussian mechanism} (Section~\ref{sec:DP_analysis}).

Note that, since Poisson quantisation is probabilistic, it might harm the accuracy of the model. Given $\mu \in s\mathbb{Z}$ and $x\in ]\mu; +\infty[$, the variance of $\quantPoissonFunc{s, \mu}(x)$ is $s^2\frac{x-\mu}{s} = s(x-\mu)$. For a small enough $s$, this variance is very small since $x$ is bounded, and there is no impact on accuracy in our experiments (Section \ref{sec:experiments}).

An important point to notice is that Poisson quantisation implies that the values to quantise have an \textit{a priori} common lower bound (otherwise the sum of the quantised values may depend on these values and not only on their sum). In our case, these values are the noised updates. The updates are already bounded by the clipping: as we will see in Section~\ref{sec:DP_analysis}, this clipping constrains the L2-norm of the updates (considered as vectors of the updates of all parameters) and thus it also constrains the absolute value of each parameter update. As for the noises, the following section shows we can consider that the noises have a common lower bound in practice.

\subsection{Dealing with unboundedness}
\label{sec:unboundedness}
\subsubsection{Bounded Gaussian noises}
The most common algorithms to sample from a Gaussian distribution are Box-Muller transform in its Cartesian and polar forms (\cite{Box-Muller,knuth97}) and the so-called ziggurat algorithm (\cite{marsaglia2000ziggurat}). As they rely on a source of uniform randomness, it is easy to see (Appendix~\ref{app:noise_bound}) that all these algorithms actually generate values whose range have bounds which are way smaller than the range of double-precision floats.
Note that the distributions followed by the outcomes of those popular sampling algorithms are almost invariably (and implicitly) considered in the literature as perfect normal distributions. We will then make the same assumption to derive our DP analysis.

\subsubsection{The problem of the unbounded Poisson distribution is not a problem}
A drawback of Poisson quantisation is that it is not bounded. Indeed, even if the Gaussian noises are bounded by the practical limitations of their sampling algorithm, the quantised noised updates are not. However, we show in this section that this is actually not a problem for our method.

A first argument to address the issue of the theoretically infinite range of the Poisson distribution would be to study Poisson sampling algorithms and find a practical bound, like we did for the normal distribution (Appendix~\ref{app:noise_bound}). Nevertheless, this does not seem that straightforward for Poisson case\footnote{Although traditional sampling algorithms generate very large size outcomes with very low probability and very large computation time.}.

Rather, let us see what happens if the Poisson sample falls out of the bounds imposed by the cryptosystem plaintext domain (the plaintext modulus). At encryption, a modulo operation will automatically be applied to the exceeding value. The same modulo operation will be performed on the aggregated updates on the server side. Observation~\ref{obs:modulo_postprocessing} shows that these two modulo operations amount to a single modulo operation on the sum of the updates, which constitutes a post-processing on this sum and, as such, does not affect the DP analysis.

\begin{observation}
    \label{obs:modulo_postprocessing}
    Let $(x_i)_{i\in \llbracket 1; K \rrbracket} \in \mathbb{Z}^K$, $N\in \mathbb{N}^*$.
    $\sum_{i=1}^K (x_i \mod N) \mod N = \sum_{i=1}^K x_i \mod N$.
\end{observation}

Recall that only integer values can be manipulated in the encrypted domain. This implies that the quantised noised updates are multiplied by the inverse quantisation scale $\frac{1}{s}$ before being encrypted, and that the participants rescale the averaged updates by $s$ once received from the server at the next round.

Let us now consider the influence of this modulo operations on the accuracy of the model. First of all, the result of the Poisson quantisation may be non-positive due to the negative offset $\mu$ and thus may fall out of $[0, \dots, N-1]$, where $N$ denotes the plaintext modulus. To avoid this situation, we make the participants send the quantised updates without adding the (potentially non-positive) offset $\mu$. When they receive the averaged updates from the server, they just have to add $\mu$ to them to get the actual averaged updates.
The second case is encountered when a sample exceeds the plaintext modulus. Nevertheless, this event is very rare if the modulus is big enough. With the parameters we use (Section~\ref{sec:experiments}), Chebyshev's inequality gives a probability lower than $1.61 \times 10^{-5}$ for this event, to compare to the number $486,654$ of parameters. In any case, our experiments prove that this has no practical influence on the model accuracy.

\subsection{DP analysis of the Gaussian mechanism}
\label{sec:DP_analysis}
According to the discussion above, our learning mechanism has the same DP guarantee as a learning mechanism where true unbounded and continuous Gaussian noise is added by the server \emph{after} aggregation, a.k.a. the Gaussian mechanism. The noise introduced as a side-effect by Poisson quantisation may even improve the privacy but, for simplicity, we consider it as banal post-processing and do not take it into account in the DP analysis.
Hence, the DP analysis reduces to the Gaussian mechanism's analysis. As explained in~\ref{sec:preliminaries_dp} and pretty much like in \cite{geyer2017differentially} for instance, we use the moments accountant \cite{abadi2016deep} to compose privacy costs in an efficient way across the multiple learning rounds.

Formally, given $\sigma\in \mathbb{R}^*_+$ the standard deviation for the aggregated noise and $S\in \mathbb{R}^*_+$ the clipping bound in L2-norm, let us consider the two density functions corresponding to two adjacent databases respectively not containing and containing the adversary's target client:
\begin{align*}
    f_1 \colon x\in \mathbb{R} \mapsto \frac{1}{\sigma\sqrt{2\pi}}e^{-\frac{x^2}{2\sigma^2}} \text{ and } f_2 \colon x\in \mathbb{R} \mapsto \frac{1-q}{\sigma\sqrt{2\pi}}e^{-\frac{x^2}{2\sigma^2}} + \frac{q}{\sigma\sqrt{2\pi}}e^{-\frac{(x-2S)^2}{2\sigma^2}}
\end{align*}
where $q = \frac{K}{M}$ is the fraction of participants by round i.e. the probability of a given client being chosen to participate in a given round. Since $q < 1$, we get privacy amplification by \emph{subsampling}.
Without loss of generality, $f_1$ is defined with mean $0$, since the integral for the moments accountants computation is on whole $\mathbb{R}$. Note that the part corresponding to the event whereby the target client is chosen as participant has an offset of $2S$ rather than $S$ because, if the absolute values are constrained by the clipping bound $S$, the actual span of the values is $2S$. As a result, changing the participant may modify the updates by $2S$.

The moments accountant of order $l\in \mathbb{R}^*_+$ corresponding to a single query to the private database, i.e. a single learning round, is:
\begin{align*}
    \alpha(l) = \max\left[\int_{\mathbb{R}}\left(\frac{f_1(x)}{f_2(x)}\right)^l f_2(x)dx,  \int_{\mathbb{R}}\left(\frac{f_2(x)}{f_1(x)}\right)^l f_2(x)dx\right].
\end{align*}

The total moments accountant of the learning process is $\alpha_{total} \leq T\max_{l\in \mathbb{R}^*_+}\alpha(l)$ where $T$ is the number of learning rounds. In practice, we compute the max for $l$ being an integer varying in $[1, \dots, 20]$.
Finally, we apply Theorem~\ref{th:tail_bound} to derive the classical DP guarantee $(\epsilon, \delta)$ from $\alpha_{total}$.

\subsubsection{Privacy cost from the point of view of a participant} For a comprehensive analysis, one must not forget that, from the point of view of a participant $k$, the noise generated by $k$ does not participate in the privatisation process\footnote{For instance, if one knows the noise that was added on a value, one just has to remove this noise from the noised value to get the initial value.}. Hence, we must take into account only the other participants' noises. The individual noises added by the participants are calibrated such that their sum has a certain standard deviation $\sigma$ i.e. these individual noises have standard deviation $\frac{\sigma}{\sqrt{K}}$, $K$ being the number of participants in each round. As a result, the DP guarantee from the point of view of a single participant must be computed by substituting $\sigma$ by $\frac{\sqrt{K}-1}{\sqrt{K}}\sigma$, which has an insignificant influence if $K$ is large ($1000$ in our experiments). Note that this is still quite conservative as it assumes that the considered participant may participate to all training rounds.

We can also interestingly extend our threat model in a straightforward way by considering that some clients may collude and share their noises with each other, quite like in~\cite{grivet2021speed}. This would result in a degraded DP guarantee, obtained by substituting $\sigma$ by $(1-\chi)\sigma$ where $\chi$ is the ratio of colluding participants.


\subsection{Homomorphic encryption protects the data (and the model) against the server}
\label{sec:HE}
While considering our learning framework protected by a distributed noising, it may not be clear why the framework ever needs to make use of cryptography. Indeed, the server receives the updates from the participants after they have been noised. However, each individual noise has been calibrated such that the \emph{aggregated} noise will obfuscate the sensitive information from a specific participant. If $\sigma$ is the standard deviation necessary to hide the data from one participant, the standard deviation of each individual noise is $\frac{\sigma}{\sqrt{K}}$. However, it should be equal to $\sigma$ if it were to protect the updates from the server \emph{before} aggregation. Such a setting is referred as \emph{local DP} in the literature. Yet, in our case, this would result in an aggregated standard deviation of $\sqrt{K}\sigma$ which would completely destroy the utility of the averaged updates, yielding a totally useless model.

In terms of concrete HE, the fact that we are considering the simple FederatedAveraging operator allows us to use additive-only schemes such as in \cite{madi2021secure} where the Paillier cryptosystem is used with batching. In the experimental results reported in the next section we have used the BFV cryptosystem which allows for more massive batching and, as such, results in much lower (amortised) overheads. Additionally, one key contribution of \cite{madi2021secure} was to associate Paillier-based homomorphic calculations to Verifiable Computing (VC) techniques (e.g. \cite{fiore2014efficiently}) to further extend the server threat model beyond the honest-but-curious one and bring execution integrity, as \cite{Madi2020} did with BFV scheme. However, these works lacked DP. Indeed, adding the DP noise on the server requires a tag that can only be generated with knowledge of the VC scheme secret key (i.e. by a client), meaning, in the Federated Learning context, that at least one of the clients would have knowledge of the noise added (resulting in a collapse of the DP guarantee for this client, even when that knowledge is uncertain). We could actually imagine that the server generates $K$ noises that sum to the required noise and send each of them to a participant for tag generation but this would double the communication cost. Hence, associating DP with VC for server-side computation integrity maintaining a reasonable communication cost requires a distributed noise generation as provided in this paper. As such, the noise generation technique provided in this work is directly applicable to setups where homomorphic calculations are paired with VC techniques.

As a very interesting side-effect, the HE layer also hides the model parameters from the server throughout the training. This may be valuable when the clients want to keep their model private, or give only a black-box access to it, either for privacy or economic reasons (cf. machine learning as a service).


\section{Experimental results}
\label{sec:experiments}
To prove the practicality of our learning framework, we performed experiments that enable us to evaluate its performance in terms of accuracy, privacy and computation time. We chose the Federated Extended MNIST (FEMNIST) dataset\footnote{Dataset available at https://www.nist.gov/itl/products-and-services/emnist-dataset} to run the experiments. The extended version of MNIST contains 62 classes (digits, upper and lower letters) of hand-written characters from 3,596 writers and comes with the writer id. FEMNIST, the federated version, was built by partitioning the data based on the writer \cite{Caldas2019}.
The network architecture is the same as in~\cite{madi2021secure}: a standard CNN composed of two convolution layers (respectively with $5*5$ kernel size and 128 channels, and with $3*3$ kernel size and 64 channels, each followed with $2*2$ max pooling), a fully connected layer with 128 units and ReLu activation, and a final softmax output layer (486,654 parameters).

\begin{figure}
    \centering
    \includegraphics[width=.5\linewidth]{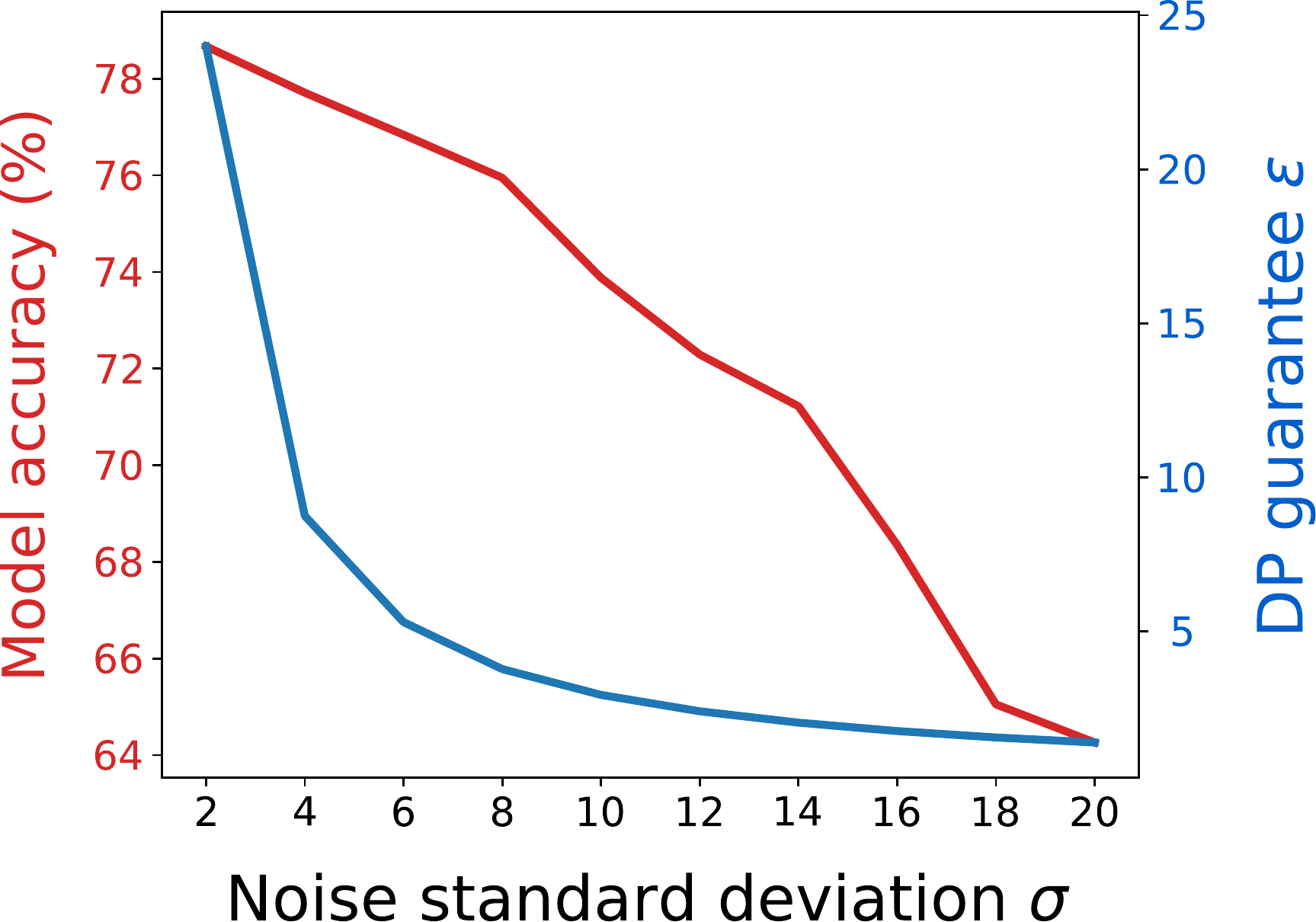}
    \caption{Model accuracy and DP guarantee vs noise standard deviation ($\delta = 10^{-5}$)}
    \label{fig:accuracy_privacy_vs_sigma}
\end{figure}

Table~\ref{tab:adaptations_accuracy} shows the influence on the model accuracy of the adaptations necessary to ensure DP. Starting from a non-DP baseline from the state of the art~\cite{madi2021secure}, we successively modified parameters of the framework, each of these modifications being required by the DP analysis\footnote{Note that the order in which we made these successive adaptations does not correspond to the order in which they are executed in the learning workflow.}.
The successive steps are:

\begin{itemize}
    \item reduce the number of learning rounds from $200$ to $100$, hence reducing the amount of queries to the clients' sensitive datasets
    \item increase the total number $M$ of clients (to 3596, the total number of writers for FEMNIST) and the number $K$ of participants per round. This has two advantages. Firstly, we can make the ratio $q=\frac{K}{M}$ smaller, decreasing the probability of a target client participating at a given round and thus the probability of this target releasing any information during this round. Secondly, the absolute value $K$ is greater, so that the information of the target participant is more diluted in the averaged updates. In practice, the experiments show that, with a fixed distortion ratio $\frac{\sigma}{K}$ fixed, which gives roughly the same model accuracy, the DP guarantee $\epsilon$ decreases when $K$ increases. We then chose $K=1000$, in our opinion the largest reasonable value so that a substantial ratio of the clients can stay idle at each round. The impact on the accuracy of the increasing of $M$ and $K$ is due to the larger number of writers, inducing a higher variety in the training samples (non-IDD across the different writers) which makes the classification task more complex.
    \item assign the same coefficient $\frac{1}{K}$ to all the participants in the weighted average (rather than the proportion $\frac{n_k}{n}$ of training samples owned by participant $k$) so that the sensitivity of the average for every participant is $\frac{S}{K}$ rather than $\max_{k\in \llbracket 1; K\rrbracket}\frac{n_k}{n}S$, where $\max_{k\in \llbracket 1; K\rrbracket}\frac{n_k}{n}$ may be much larger than $\frac{1}{K}$
    \item clip updates with clipping bound $S$ to ensure finite sensitivity (we took $S=1$ which has a mild impact on privacy and allows for good DP guarantees)
    \item on the participant side, quantify the noised updates via Poisson quantisation
    \item apply a modulo operation on the noised updates on the participant side, and on their sum on the server side, which is automatically done by encryption
    \item add the Gaussian noise necessary to make the learning process differentially private. We chose $\sigma=6$ for the total noise because it gives a good trade-off between privacy and model accuracy as shown in Figure~\ref{fig:accuracy_privacy_vs_sigma}.
\end{itemize}

We fixed the scale for Poisson quantisation to $10^{-4}$, as in~\cite{madi2021secure}, since it does not much affect the accuracy.
We used as common lower bound $\mu$ of the Gaussian noises the lower bound from the ziggurat algorithm with $255$ rectangles, i.e. $15.81$ (see Appendix~\ref{app:noise_bound}), of course multiplied by the standard deviation of the distribution. This lower bound is greater (in absolute value) and then more conservative than the lower bounds of the two other sampling algorithms we considered. Moreover and quite importantly, ziggurat algorithm with 255 rectangles is actually the one chosen by the numpy library we used.

The whole training process is $(\epsilon, \delta)$-differentially private, with $\epsilon = 5.31$ and $\delta = 10^{-5}$.
Actually, for $\delta=10^{-5}$, $\epsilon=5.306$ for an end-user which is not a participant and $\epsilon=5.313$ for a participant (see discussion at the end of Section~\ref{sec:DP_analysis}). Together with the model accuracy of $76.84\%$ we get the same privacy/utility tradeoff as~\cite{agarwal2021skellam,kairouz2021distributed} got with secure aggregation. Nevertheless, if communication is a critical issue, we may use a greater quantisation scale, at the expense of accuracy, but this would not harm the DP guarantee, contrary to \cite{agarwal2021skellam,kairouz2021distributed}.
Interestingly, we experimentally notice that the quantisation and modulo operation have no influence on the accuracy: the model trained with noise but without quantisation or modulo operation still has an accuracy of $76.84\%$.

\begin{table}[!htb]
    \begin{minipage}[t]{.4\linewidth}
        \centering
        \caption{Influence of successive adaptations on accuracy.}
        \label{tab:adaptations_accuracy}
        \begin{tabular}{ l | c }
            & Accuracy \\
            \hline
            State of the art~\cite{madi2021secure} & $84.6\%$ \\
            \hline
            Decrease the number \\of learning rounds $T$ & $83.58\%$ \\
            \hline
            Increase $M$ and $K$ & $81.04\%$ \\
            \hline
            Assign same coefficients & $80.21\%$ \\
            \hline
            Clipping of the updates & $79.26\%$ \\
            \hline
            Quantisation & $79.03\%$ \\
            \hline
            Modulo operation & $79.07\%$ \\
            \hline
            Adding random noise & \textbf{76.84\%} \\
        \end{tabular}
    \end{minipage}
    \hspace{3em}
    \begin{minipage}[t]{.4\linewidth}
        \centering
        \caption{Computation time (in seconds) of HE operations with a 26 bits modulus for the \emph{full} 486654 weights model.}
        \label{tab:HE_time}
        \begin{tabular}{ l | c }
            Number of\\participants $K$ & $1000$ \\
            \hline
            Context and\\key generation & $0,05698$ \\
            \hline
            Encoding & $0,05704$ \\
            \hline
            Encryption & $0,84642$ \\
            \hline
            Evaluation & $26,22508$ \\
            \hline
            Decryption & $0,29308$ \\
        \end{tabular}
    \end{minipage}
\end{table}

The experimental results for HE were realised with BFV in batched mode and Palisade library (version 1.1.6) on an Intel Core i7 with 4 cores at 3 GHz with 32 GB Ram on Ubuntu 18.04. The security level was set to 128 bits and the batch size used was of 8,192. Following this the overall 486,654 updates can be packed in only 60 ciphertexts (where each of the 8,192 slots contains one gradient update). Table \ref{tab:HE_time} provides the overall homomorphic computation time \emph{for the full model} for a $26$ bits modulus and 1000 participants per round resulting in a (fairly practical) maximum of 26 seconds of homomorphic calculations (per FL round). Performing the full FL cycle (without communications) on a GPU-based HPC cluster takes around 20 hours (i.e., 12 minutes per FL round), resulting in a 3.6\% computation time overhead imputable to HE calculations.

The choice of a $26$ bits modulus is due to an empirical investigation. For $26$ bits or more, the model trains correctly, with almost no impact of the modulo operation on the accuracy (see Table~\ref{tab:adaptations_accuracy}). Below $26$ bits, the model does not learn at all. This sharp change of behaviour is due to the fact that the modulus exponentially depends on the number of bits and that the distribution of the quantised noised updates is actually very peaked - the ratio standard deviation over expectation is lower than $2.22 \times 10^{-3}$.


\section{Conclusion and perspectives}
\label{sec:conclusion}
In this work, we addressed the problem of collaborative learning where all the actors of the training stage - clients and server - and the end-users are potential honest-but-curious adversaries. We chose the popular FL framework and adapted it in two ways. Firstly, by having the clients add random noise on the information they send to the server, we made the learning mechanism differentially private from the point of view of any end-user of the model and that of the clients themselves. Secondly, following~\cite{madi2021secure}, we added a HE layer on the server side so that the server cannot see the updates coming from the clients, that may release sensitive information about the training data.
The HE layer has a major impact on the random noise added to ensure DP, essentially because of the limited number of bits available for a reasonable computation time. However, we proved that this interference can be seamlessly dealt with in terms of privacy thanks to some adaptations among which a new carefully crafted quantisation operator.

We ran experiments on FEMNIST dataset that prove the practicality of our framework in terms of accuracy, privacy cost and computation time, thoroughly analysing the cost of DP in accuracy compared to an non-DP baseline.

The present work could easily be extended to a setting whereby the assumption of the honest-but-curious server is relaxed, making the learning process robust to a server who would, willingly or not, make mistakes in its computations. As argued in Section \ref{sec:HE}, this could be done using verifiable computing techniques, as in~\cite{madi2021secure}, in a quite straightforward further work thanks to the fact that the server is not in charge of adding the random noise necessary to DP.

Testing our framework on a larger, more cross-device-oriented dataset would be quite interesting to estimate its scalability. Moreover, this could be advantageous on the privacy point of view since this would allow to increase the number of clients $M$ and thus having simultaneously a big number of participants $K$ and a low ratio $\frac{K}{M}$, conditions that will both improve the DP guarantee.

Another quantisation function or a more involved analysis that would not need to lower bound the random noise added to the updates would allow us to get rid of the argument of the imperfect sampling algorithms and to use our framework with other noise distributions, not necessarily bounded, even in practice.


\section{Ethical principles}

The legal landscape in Artificial Intelligence is evolving quickly, following growing public concerns about weaknesses, risks, and misuse of artificial intelligence (AI). In Europe, United States and China, laws and regulations are evolving to address these concerns. In 2016, the European Union provided an answer to the protection of individual’s privacy against information leakage from a legal standpoint by publishing the General Data Protection Regulation – GDPR. It formally provides with the following key principles:
\begin{itemize}
    \item \emph{Transparency}: Personal data must be processed with transparency, and data owners can ask to be informed at any time on how their data are being used.
    \item \emph{Limited purposes and retention}: Personal data must be collected for clear, understandable and legitimate purposes, and should not be processed for any other purpose than the initial ones. Moreover, data must be kept for a period that does not exceed the processing of the selected purposes.
    \item \emph{Data integrity and privacy}: Personal data must be processed in such a way as to ensure appropriate security of such data, including protection against unauthorised or unlawful processing and against accidental loss.
    \item \emph{Access, rectification, erasing}: Data owners should be granted easy access to their personal data as well as the possibility to rectify of any inaccuracy or to erase any personal data.
    \item \emph{Privacy by design}: Each new technology or application processing personal data, or making it possible to process it, must ensure, from its design and each time it is used, that it incorporates all the protection principles of the GDPR.
\end{itemize}
Our paper aims at reconciling artificial intelligence practices with existing and future regulations on privacy requirements. It empowers citizens with innovative tools to ensure personal data privacy. Therefore we argue that this work is not only compliant with current ethical expectations, but also contributes to improve future legal and ethical requirements related to artificial intelligence.


\bibliographystyle{splncs04}
\bibliography{./bibliography/mybiblioFL,./bibliography/biblioDP,./bibliography/biblioAttacks, ./bibliography/biblioCrypto}

\newpage
\appendix

\section{Sampling algorithms and lower bound of the Gaussian noise}
\label{app:noise_bound}
As mentioned in~\ref{sec:unboundedness}, we consider the three most popular algorithms for sampling Gaussians and show that the samples belong to a (small) bounded range.

\textit{\textbf{Box-Muller transform (Cartesian form) \cite{Box-Muller}:}}
The Cartesian form of the Box-Muller transform samples two independent uniform random variables $U_1$ and $U_2$ in $[0; 1]$. The random variables $Z_1 = \sqrt{-2\log(U_1)}\cos(2\pi U_2)$ and $Z_2 = \sqrt{-2\log(U_1)}\sin(2\pi U_2)$ are independent and follow a standard normal distribution (standard deviation $1$ and mean $0$).
Since the function $\cos$ and $\sin$ are bounded by $-1$ and $1$, we see that the maximum absolute value of $Z_1$ and $Z_2$ is reached for the minimum value of $U_1$ which is $2^{-\nBits}$ where $\nBits$ is the number of bits used to represent an integer. For $64$ bits, we get $\sqrt{-2\log(2^{-64})} \approx 9.42$.

\textit{\textbf{Box-Muller transform (polar form) \cite{knuth97}:}}
The polar form of Box-Muller transform samples two independent uniform random variables $U_1$ and $U_2$ in $[-1; 1]$ and calculates $s = U_1^2 + U_2^2$. The random variables $Z_1 = \sqrt{-2\log(s)}\frac{U_1}{\sqrt{s}}$ and $Z_2 = \sqrt{-2\log(s)}\frac{U_2}{\sqrt{s}}$ are independent and follow a standard normal distribution.
In this case, $\frac{U_1}{\sqrt{s}}$ and $\frac{U_2}{\sqrt{s}}$ always belong to $[-1; 1]$ hence the maximum absolute value reached by $Z_1$ and $Z_2$ is $\sqrt{-2\log(s_{min})}$ where $s_{min}$ is the minimum value possibly reached by $s$, namely $s_{min} = (2^{-\nBits})^2 + (2^{-\nBits})^2 = 2^{-2\nBits + 1}$.
For a $64$ bits processor, this gives $\sqrt{-2\log(2^{-127})} \approx 13.27$.

\textit{\textbf{Ziggurat algorithm \cite{marsaglia2000ziggurat}:}}
The ziggurat algorithm applies to monotonically decreasing probability distributions and extends to symmetric unimodal distributions like the normal one by randomly choosing on which side of the mode the sampled value will fall. The algorithm works by covering the distribution by stacked rectangular regions of same area. What matters for our problem of finding the bound of the sampling process is only the tail rectangle. In the rare case where the value did not fall into one of the other rectangles, a fallback algorithm is used to sample the value from the tail.

Let $x_{tail}$ be the abscissa of the right side of the last rectangular region before the tail.
The fallback algorithm for the normal distribution samples two independent uniform random variables $U_1$ and $U_2$ from $[0;1]$ and defines $x=-\log(U_1)/x_{tail}$, $y=-\log(U_2)$. It then tests if $2y > x^2$ and returns $x + x_{tail}$ if yes. Otherwise, it restarts with two new samples $U_1$ and $U_2$.
Thus, the biggest value in the monotonic case, also the biggest absolute value in the symmetric case, is $-\log(2^{-\nBits})/x_{tail} + x_{tail}$. In \cite{marsaglia2000ziggurat}, Marsaglia and Tsang calculate that, for $255$ rectangles, $x_{tail} \approx 3.65$. For $64$ bits, we then get the bound $15.81$.

These "artificial" bounds, that we cannot avoid in practice anyway, are justified by the very low probability of a draw outside them: less than $10^{-20}$ for the lowest bound, $9.42$, and less than $10^{-55}$ for the highest, $15.81$.
To get a sample from an arbitrary normal distribution, it suffices to scale the sample of the standard normal distribution by the wanted standard deviation and then add the wanted mean.
This discussion allows us to exhibit a lower bound for the Gaussian noises and, the unnoised updates being bounded by the clipping, for the noised updates. As a consequence we can apply Poisson quantisation.

\end{document}